# MeerKAT HI observations of Low Surface Brightness/Ultradiffuse Galaxy Candidates Projected around Two Southern Loose Groups

Chandreyee Sengupta,[1,2] Tom C. Scott,[3] Hao Chen,[4] Hyein Yoon,[5,6,7] Yogesh Chandola,[2] Mengtian Li,[8,9] Gyula I. G. Józsa,[10,11] O. Ivy Wong,[12,13] Yin-Zhe Ma,[14] Patricio Lagos,[15,3] Ruta Kale,[16] and Denis Tramonte[17]

[1]*Centre for Space Research, North-West University, Potchefstroom 2520, South Africa*
[2]*Purple Mountain Observatory (CAS), No. 10 Yuanhua Road, Qixia District, Nanjing 210023, China*
[3]*Institute of Astrophysics and Space Sciences (IA), Rua das Estrelas, 4150–762 Porto, Portugal*
[4]*Research Center for Astronomical Computing, Zhejiang Laboratory, Hangzhou 311100, China*
[5]*Institute for Data Innovation in Science, Seoul National University, 1 Gwanak-ro, Gwanak-gu, Seoul 08826, Korea*
[6]*Astronomy Program, Department of Physics and Astronomy, Seoul National University, 1 Gwanak-ro, Gwanak-gu, Seoul 08826, Korea*
[7]*Sydney Institute for Astronomy, School of Physics A28, The University of Sydney, NSW 2006, Australia*
[8]*Key Laboratory of Radio Astronomy, Purple Mountain Observatory, Chinese Academy of Sciences, Nanjing 210023, China*
[9]*School of Astronomy and Space Science, University of Science and Technology of China, Hefei, Anhui 230026, China*
[10]*Max-Planck-Intitut für Radioastronomie, Auf dem Hügel 69, 53121 Bonn*
[11]*Department of Physics and Electronics, Rhodes University, PO Box 94, Makhanda 6140, South Africa*
[12]*CSIRO Space & Astronomy, PO Box 1130, Bentley, WA 6102,Australia*
[13]*International Centre for Radio Astronomy Research-M468,The University of Western Australia,Crawley, WA 6009, Australia*
[14]*Department of Physics, Stellenbosch University, Matieland 7602, South Africa*
[15]*Institute of Astrophysics, Facultad de Ciencias Exactas, Universidad Andrés Bello, Sede Concepción, Talcahuano, Chile*
[16]*National Centre for Radio Astrophysics (NCRA), Tata Institute of Fundamental Research (TIFR), Pune 411007, India*
[17]*Department of Physics, Xi'an Jiaotong-Liverpool University, 111 Ren'ai Road, Suzhou Dushu Lake Science and Education Innovation District, Suzhou Industrial Park, Suzhou 215123, People's Republic of China*

## ABSTRACT

A large catalogue of low surface brightness galaxies (LSBGs) from the Dark Energy Survey showed significant clustering around nearby galaxy groups and clusters. Using the HIPASS survey, we tried to determine the redshift of a sub–sample of these LSBGs and determine whether they were members of the groups they were projected near, but this was hampered by HIPASS's high spectral rms. This letter reports on MeerKAT H I observations to determine the redshifts of 52 LSBG candidates projected in the vicinity of two groups from our previous HIPASS study. The main goal is to investigate and ascertain whether these LSBGs are genuine group members. H I was detected with MeerKAT and redshifts were determined for only five of the 52 candidates within a velocity range of ± 2500 km s$^{-1}$ of their respective group velocities. All five H I detections were blue LSBGs and two of them were confirmed to be ultradiffuse galaxies (UDGs). Both these UDGs were group members, while the other three detections were either foreground or background galaxies. In this letter we explore scenarios that can explain the 90% non-detection. MeerKAT's excellent sensitivity allows us to conclude that the majority of the non–detected candidates, particularly the blue galaxies, are not group members but lie at higher redshifts. However, this still leaves the open question as why Tanoglidis LSBG candidates, in particular the red ones, appear to be clustered in projection around nearby groups.

*Keywords:* galaxies: distances and redshifts — galaxies: groups: individual (NGC1200, NGC7396) — galaxies: ISM — galaxies: photometry

## 1. INTRODUCTION

Using optical imaging from the Dark Energy Survey (DES; Abbott et al. 2018), Tanoglidis et al. (2021) reported ∼23790 LSBG candidates in an area of ∼ 5000 deg$^2$. According to their definition, these are galaxies with g – band effective radii ($R_\mathrm{e}$) ≥ 2.5″ and

Email: sengupta.chandreyee@gmail.com

mean surface brightness $\mu_{eff(g)} \geq 24.2$ mag arcsec$^{-2}$. This opened up an excellent opportunity to study the low surface brightness and low mass end of the galaxy mass function. Although Tanoglidis' LSBG candidates are distributed across the southern sky, a large fraction of them were found to be clustered in projection around prominent nearby galaxy groups and clusters. This clustering effect was significantly stronger for red LSBG candidates than for blue candidates. Making the assumption that the clustering of LSBG candidates around these clusters/galaxy groups is also true in velocity space, and not just in projection, Tanoglidis et al. (2021) reported 80 such groupings of candidates. Making the same redshift assumption we classified a fraction of the Tanoglidis et al. (2021) LSBG candidates as UDGs as defined by van Dokkum et al. (2015), using the criteria of a g – band $R_e \geq 1.5$ kpc and the central surface brightness $\mu_{o(g)} \geq 24.0$ mag arcsec$^{-2}$. UDGs were first reported in and surrounding the Coma cluster (van Dokkum et al. 2015), and were found to be relatively more abundant in massive clusters than in groups (van der Burg et al. 2017). As a result the majority of the UDG literature has been focused on UDGs in clusters or groups.

Using archival HIPASS data, Zhou et al. (2022) searched for H I counterparts in 22 nearby Tanoglidis et al. (2021) groups containing a total of 409 LSBG candidates. The search resulted in H I non–detections for all 409 target LSBGs. Zhou et al. (2022) concluded that while the main impediment of their study was the high spectral rms of the HIPASS survey, the high rate of non–detections was not solely attributable to the high rms. Even with the high rms limitation, the results indicated a strong possibility that a good fraction of the LSBGs and UDG candidates were distant background galaxies and not part of a Tanoglidis group. This claim has serious implications. Therefore, it is crucial to follow up the HIPASS study with more sensitive H I observations. In this letter, we present MeerKAT H I observations of two groups from our HIPASS sample of 22 nearby southern groups in Zhou et al. (2022). The aim of this letter is to investigate and ascertain whether the projected clustering of Tanoglidis LSBGs around the two groups is also true in velocity. In other words, whether the Tanoglidis LSBGs are genuine group members. Using MeerKAT significantly improves the signal to noise (SNR) compared to HIPASS (Meyer et al. 2004), allowing us to draw more robust conclusions about the commonly accepted clustering properties of optically identified UDGs.

The layout of the paper is as follows. In Section 2 we give a description of the two groups studied in this work. In Section 3 we discuss the MeerKAT observations and data reduction. Section 4 presents the results, i.e., the H I detections and their properties. Section 5 discusses the scientific implications of the results and finally Section 6 summarises our main results. All RA ($\alpha$) and declination ($\delta$) positions referred to throughout this letter are J2000 coordinates.

## 2. SAMPLE GROUPS

We selected two of the nearest Tanoglidis et al. (2021) galaxy groups from Zhou et al. (2022) to be studied with MeerKAT. These two groups had 52 LSBGs of which 4 were UDGs candidates (Figure 1). The groups, NGC 1200 and NGC 7396, are at distances of 57 and 68 Mpc, respectively (Tanoglidis et al. 2021).

Díaz-Giménez & Zandivarez (2015) lists the NGC 1200 group (v = 3937±145 km s$^{-1}$) in their catalogue of galaxy groups (ID 425) as having 6 members, including NGC 1200. Additionally, Xu et al. (2018) reported the group as having X–ray emitting intra-group medium (IGM) with a $r_{500}$ extent of 30'. Alternatively, the NGC 7396 group has been reported as a low density contrast group (LDCE 1541) (V=4809±124.2 km s$^{-1}$) (Crook et al. 2007) with 4 members.

After the initial data release, Tanoglidis et al. (2021) published an updated version of their data tables. A revised data table added two and removed two LSBG/UDG candidates from the projected region surrounding each of our groups. Additionally, in the revised table, LSBGs previously associated to the NGC 7396 group were considered as field galaxies, and not associated to the group. In our analysis we consider all LSBG/UDGs appearing in either data release as group candidates. The NGC 7396 and NGC 1200 groups have 20 and 32 LSBG candidates, respectively. Figure 1 shows the positions of the Tanoglidis LSBG/UDG candidates projected within each MeerKAT pointing.

## 3. OBSERVATIONS AND DATA REDUCTION

H I 21 cm line observations were carried out for two galaxy groups, NGC 1200 and NGC 7396, using the SKA precursor telescope MeerKAT from May to September 2021. The MeerKAT full width half maximum (FWHM) primary beam is $\sim 1°$ at 1420.4057 MHz. Two MeerKAT slightly overlapping pointings (P1 and P2) were made for each group to cover all LSBG/UDG candidates associated with each Tanoglidis et al. (2021) group, see Figure 1. The total on-source integration time for each NGC 1200 group pointing was 3.8 hours, and that for each of NGC 7396 group pointing was 2 hours.

MeerKAT offers wideband data and in L–band ranging from 900 MHz to 1670 MHz. The main aim of this work was to establish whether the LSBG/UDG candidates projected near each group contained H I at the group redshift, which is one way of ascertaining if they were group members. We therefore restricted the results presented here to a spectral window centered at each group's systemic velocity ±2500 km s$^{-1}$. While a ± 250 km s$^{-1}$ velocity range would have sufficed to determine whether H I detected LSBG/UDGs were group members, we used a larger (approximately ±2500 km s$^{-1}$) velocity range to determine if some of the LSBG/UDGs candidates were located in the background or foreground. The observations were carried out using the 32k correlator mode, with a velocity resolution of 5.5 km s$^{-1}$.

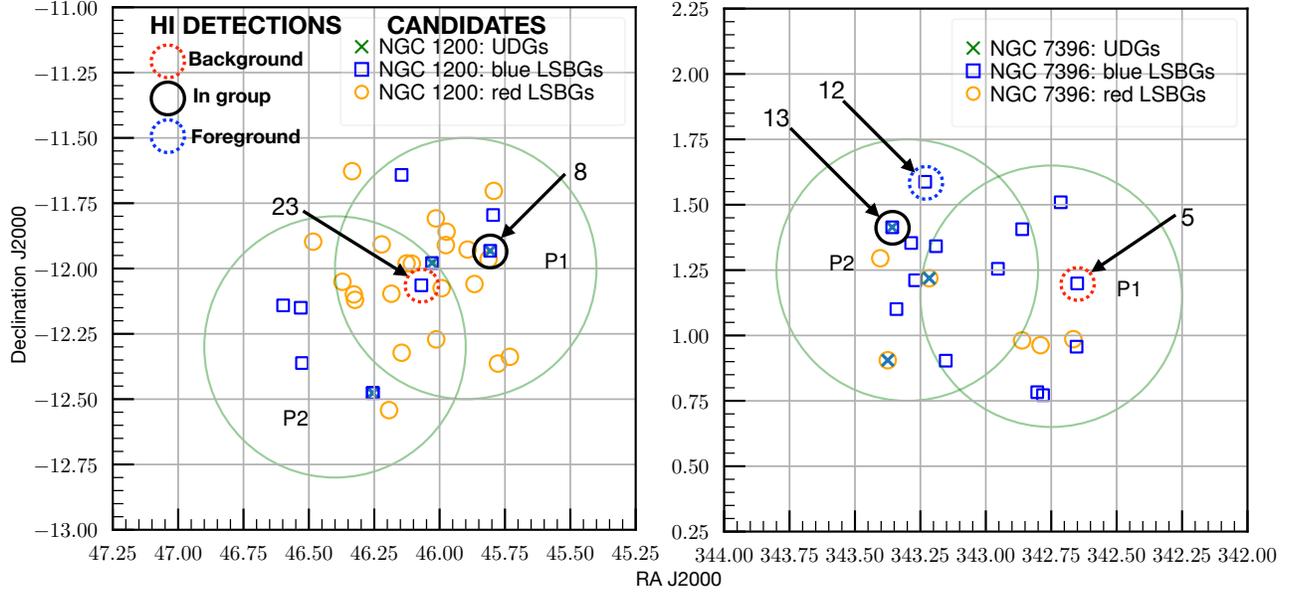

**Figure 1.** 1° FWHP primary beams for the MeerKAT pointings are shown with the large green circles and marked P1 and P2 for NGC 1200 (left) and NGC 7396 (right). The projected positions of the LSBG and UDG candidates and their DES g–z colors are indicated by the symbols in the legend. Candidates with an H I detection are shown with mid sized circles and the candidate's ID number. The circle's colour indicates the H I detection's velocity relative to its group per the legend. Note: Two LSBGs in NGC 1200 group, both candidate "blue UDG" have very similar coordinates (46.25, -12.47) and thus appear as one galaxy in Figure 1. The group centre coordinates are the following: NGC 1200 45.97, -11.99 and NGC 7396 343.09, +1.09

The initial MeerKAT data reduction was carried out using Containerized Automated Radio Astronomy Calibration (CARACAL) software (Józsa et al. 2020) for flagging, gain and bandpass calibrations. The gain calibrated data were then self calibrated using the Common Astronomy Software Applications package (CASA) TASKS GAINCAL, APPLYCAL and TCLEAN (CASA Team et al. 2022) iteratively on the data, until the calibration was robust and no further improvement in the map rms or SNR were detectable. The self–calibrated data was then continuum subtracted using first UVSUB and then UVCONTSUB, and imaged using TCLEAN with natural weighting and with a taper to produce 20″ synthesised beams. Finally, the integrated H I maps (moment maps) were made using IMMOMENTS. Each group's candidates are projected across the entire primary beam areas. Thus the primary beam correction routine (Mauch et al. 2020) was applied to the final H I cubes. According to (Tanoglidis et al. 2021), 52 newly discovered LSBGs are clustered in projection around the NGC 1200 and NGC 7396 groups. The aim of this letter was to search for H I along the 52 lines of sight and ascertain whether the LSBGs belonged to the groups. Spectra were extracted from the H I cubes at the positions of the LSBG candidates using the casa task SPECFLUX within a 1′ × 1′ box. At the distance of the groups, this samples ∼ 16 to 19 kpc$^2$, enough to accommodate small dwarf like galaxies with an H I envelope. We detected H I in five LSBGs. Additionally, we carried out SoFiA 2 test runs with different signal to noise thresholds but that did not yield any additional significant signals. SoFiA 2 returned the same results as our manual spectra extraction method and confirmed our detections.

## 4. RESULTS

Searching within approximately ±2500 km s$^{-1}$ of their Tanoglidis group's velocity for each of the 52 lines of sight associated with the candidates, we detected H I at only 5 positions. These were at positions 8 and 23 for NGC 1200 and positions 5, 12 and 13 for NGC 7396 (Figure 1). The numbers indicate the group members in our group membership tables which are not presented in this letter. For their coordinate information, please see Table 1. For the positions where H I was not detected, stacking of the P1 and P2 spectra did not yield any additional detections.

Figure 2 shows the integrated H I contours overlaid on DES (DESI Collaboration et al. 2023) g–band images of the five candidates detected in H I. With a 3.8 hours on–source integration time each pointing for NGC 1200 group, the per channel rms reached after primary beam correction was ∼ 0.26 mJy beam$^{-1}$. For NGC 7396 the on–source integration time per pointing was 2.0 hours and the rms reached was ∼ 0.45 mJy beam$^{-1}$. After primary beam correction the map rms changes with distance from the centre and is highest at the edges of the map. The rms values given here are representative values for the entire cubes, approximating the median of several measurements along the primary beam radius. Assuming a genuine H I emission signal in three consecutive channels, the three sigma H I mass sensitivity reached at the distances of the NGC 1200 and NGC 7396 groups are 1.0×$10^7$ and 2.4×$10^7$ M$_\odot$, respectively. All the five MeerKAT H I detections were Tanoglidis blue LSBG candidates. Our H I study revealed candidate 23 from the NGC 1200 group and candidate 5 from the NGC 7396 as background galaxies while candidate 12 from the NGC 7396 group is a foreground galaxy. Only candidate 8 from the NGC 1200 group and 13 from the NGC 7396 group were confirmed as group members (see Table 1 for the velocity information). Additionally, amongst the H I detections, these two candidates, both blue LSBGs, also qualified as UDGs using the surface brightness and effective radii estimates from van Dokkum et al. (2015). Our H I detections provided the spectroscopic confirmation of their distances which were critical for validating them as genuine UDGs. Table 1 summarises the properties of the H I detected LSBGs and lists the coordinates, g–i colour, g–band central surface brightness, g–band effective radii estimated using distances from their H I detections, LSBG/ UDG category, group membership, H I velocity, W$_{20}$ of the H I spectrum and the H I mass of the galaxy.

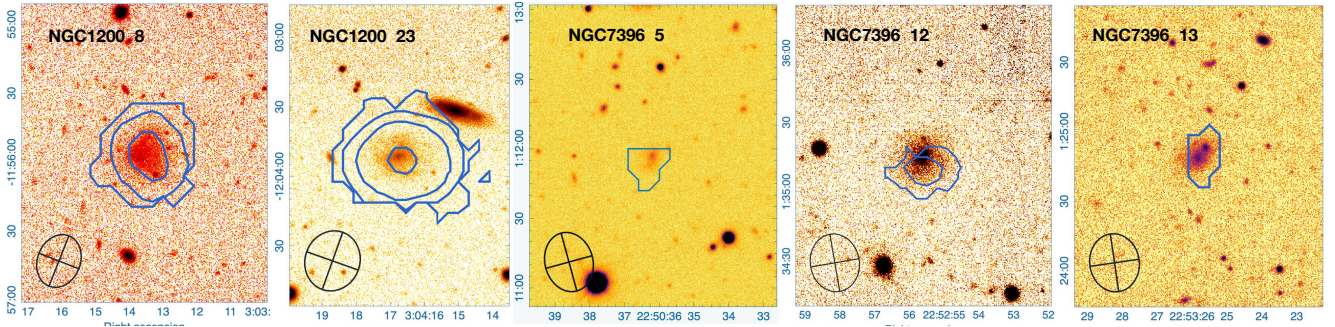

**Figure 2.** Integrated H I contours overlaid on DES g–band images of the five H I detected UDG candidates, with the galaxy group indicated at the top of each panel followed by its candidate number. The column densities of the contours are as follows (from left to right): image(1) 1.7, 2.3, 16.0 ×$10^{19}$ cm$^{-2}$, image(2) 1.8, 4.9, 7.7, 14.3 ×$10^{19}$ cm$^{-2}$, image(3) 2.2 ×$10^{19}$ cm$^{-2}$, image(4) 2.8, 3.7 ×$10^{19}$ cm$^{-2}$, image(5) 1.8 ×$10^{19}$ cm$^{-2}$. The ellipse at the bottom left of each panel shows the size and position angle of the MeerKAT synthesised beam.

## 5. DISCUSSION

Compared to our previous HIPASS H I study of LSBG/UDG candidates in southern groups (Zhou et al. 2022), our MeerKAT observations were an order of magnitude more sensitive. However, even with MeerKAT's improved sensitivity, 90% of the LSBG/UDG candidates remained H I non–detections. In both of our groups, the target galaxies had a mix of DES g − i colours, i.e., red (g − i > 0.6 mag) and blue (g − i ≤ 0.6 mag). In the NGC 1200 group, 10 out 32 LSBG/UDG candidates and in NGC 7396 group, 14 out of 20 LSBG/UDG candidates had a g–i blue colour. Our 10% MeerKAT H I detection rate, all blue galaxies, was lower than the 25% rate reported for UDGs projected in or around the Coma cluster in Karunakaran et al. (2020). Their H I detections were also predominantly blue galaxies. Although, their H I detections were either in the background or foreground of the cluster and the majority of candidates irrespective of colour were not detected in H I. That the majority of our LSBG/UDGs, and particularly the red ones, were undetected in H I suggests two possibilities:

1) The LSBG/UDGs are in fact members of the groups, but the high rate of non–detection is due to the candidates being severely H I deficient. While this is a credible explanation in the intra–cluster medium rich clusters, e.g., Coma, the much lower velocity dispersion in groups (≲ 250 km s$^{-1}$) combined with the lower intra–group medium (IGM) densities makes this less likely in our groups. However, this could be one of the possible reasons for the H I non-detections in the red LSBG dominated NGC 1200 group whose IGM X–ray luminosity is 2.82±0.086 × $10^{42}$ ergs s$^{-1}$ (Xu et al. 2018), two orders of magnitude lower than in the A 1367 cluster (Plionis et al. 2009) where ram pressure stripping (RPS) is observed (Scott et al. 2018). For the NGC 7396 group, a blue LSBG dominated loose group without IGM X–ray emission, this explanation seems even less likely. Normally, H I rich dwarfs and even H I deficient dwarfs often form a significant population in loose groups (Sengupta & Balasubramanyam 2006; Pisano et al. 2011). Following the H I deficiency estimation method in (Haynes & Giovanelli 1984), field dwarf galaxies, with optical diameters between 5 to 10 kpc, are expected to contain ∼ 1.8 × $10^8$ M$_\odot$ to 7.4 × $10^8$ M$_\odot$ of H I. A dwarf galaxy is considered H I deficient if it has a significantly lower H I content compared to the average H I content of similar field dwarf galaxies. For our groups, the H I mass sensitivity reached is an order of magnitude lower than the expected H I content of lower mass field dwarf galaxies. Thus for any dwarf galaxy member of our groups, an H I non–detection implies it is severely H I deficient. At our groups' distance, even if some or most LSBGs were H I deficient, the excellent sensitivity levels reached in this experiment should have allowed us to detect H I in more partially H I stripped galaxies, especially those with a blue colour.The fact that we do not detect them, leads us to conclude that they are unlikely to be group members.

2) The candidates appear in projection to be group members, but in reality they are most likely background galaxies. Among our 5 detections, only two are unambiguously group members, with their radial velocities within ± 200 km s$^{-1}$ of their group velocities. Candidate 23 in the NGC 1200 group is a background galaxy and candidates 5 and 12 of NGC 7396 group are background and foreground galaxies respectively. Based on our highly sensitive H I observation results, we conclude that for NGC 1200 and NGC 7396, the H I non–detections of the LSBGs and UDGs identified in the Tanoglidis et al. (2021) as potential group members are likely distant background galaxies.

It could be argued that these are undetected foreground galaxies. In this work, we explored with MeerKAT, for the NGC 1200 group velocities ≥ 1437km s$^{-1}$ (∼ 20 Mpc) and NGC 7396 ≥ 2309 km s$^{-1}$ (∼ 32 Mpc). Mass sensitivities reached with MeerKAT at the lowest velocity edge of our sampled spectral window, at distances 20 Mpc for NGC 1200 group's pointings and at distance 32 Mpc for NGC 7396 group's pointings are similar to 1.2× 10$^6$ M$_\odot$ and 5.3× 10$^6$ M$_\odot$ respectively. This is almost an order of magnitude higher sensitivity that we quote for the actual group distances. Typical dwarf galaxies, even the H I deficient ones, would have been detected if they were present at or above these distances. In our previous work, we searched for H I counterparts for Tanoglidis LSBG candidates with the HIPASS data (Zhou et al. 2022) in 18 groups, including NGC 1200 and NGC 7396. Those searches extended to velocities below those explored with MeerKAT, but did not reveal a single LSBG candidate H I counterpart. The HIPASS mass sensitivities at low redshifts, e.g., 10 Mpc and 20 Mpc are approximately $5 \times 10^7$ M$_\odot$ and $2 \times 10^8$ M$_\odot$, respectively. These mass sensitivities are lower than the average H I content of normal dwarfs. So despite the low sensitivity, we would have expected to detect H I in a few LSBG candidates with HIPASS had they been present in the low velocities, i.e. in the foreground. The fact that we didn't detect any, lead us to hypothesise that a majority of non H I detected Tanoglidis LSBG candidates in our HIPASS and MeerKAT studies are most likely distant background galaxies.

It is possible that these LSBGs are parts of large-scale filamentary structures and hence we see the projected alignment as clustering of LSBGs (Palumbo et al. 1993; Blue Bird et al. 2020; Hoosain et al. 2024). However, in the 5000 km s$^{-1}$ velocity range (group velocity ± 2500 km s$^{-1}$ ) that we investigated in this work, we did not detect a sufficient number of galaxies to support this argument. But, one cannot rule out the possibility of filamentary structures beyond this velocity range. A preliminary investigation into this possibility did not reveal evidence of such structures. A full investigation of this possibility is beyond the scope of this letter. Future multiwavelength studies could reveal more information about the presence of large-scale structures incorporating the NGC 1200 and NGC 7396 groups.

## 6. CONCLUSIONS

Using MeerKAT H I observations we confirmed the redshifts of 5 Tangolidis LSBGs candidates. All 5 are blue LSBGs and only 2 are group members. And both these group members were confirmed as UDGs. None of the 28 red Tangolidis LSBG/UDG candidates in either group were detected in H I. Based on our MeeerKAT H I results we hypothesise that the majority of the Tanoglidis LSGB/UDG candidates projected near the NGC 1200 and NGC 7396 groups are most likely background galaxies with velocities > 2500km s$^{-1}$ above their respective group's velocity. This result is consistent with H I studies of UDG candidates around clusters which also predominantly report H I non–detection of UDGs (Karunakaran et al. 2020; For et al. 2023). However, we note that gas depletion processes and timescales are expected to be different in clusters compared to loose groups. For example, ram pressure removal is expected to be significantly less efficient in groups than in clusters. A possibility could be that these LSBGs are parts of a background large scale structure. A more detailed investigation is needed to confirm that. This still leaves the open question of why the red LSBGs appear to be more clustered around the groups compared to the bluer galaxies.

## ACKNOWLEDGMENTS

The MeerKAT telescope is operated by the South African Radio Astronomy Observatory, which is a facility of the National Research Foundation, an agency of the Department of Science and Innovation. We thank the referee for his/her constructive comments and suggestions to improve the letter. We thank Catarina Lobo for her advice and discussions. PL (DOI 10.54499/dl57/2016/CP1364/CT0010) and TS (DOI 10.54499/DL57/2016/CP1364/CT0009) are supported by national funds through Fundação para a Ciência e a Tecnologia (FCT) and the Centro de Astrofísica da Universidade do Porto (CAUP). HC was supported by National Key R&D Program of China NO. 2023YFE0110500, and by the Leading Innovation and Entrepreneurship Team of Zhejiang Province of China (Grant No. 2023R01008). HY was supported by the National Research Foundation of Korea (NRF) grant funded by the Korea government (MSIT) (RS-2025-00516062). YC thanks the Center for Astronomical

Table 1. MeerKAT LSBG/UDG H I detections

| Group | Candidate ID [c] | RA(J2000) h:m:s | Dec(J2000) d:m:s | g–i mag | $\mu_o$ [a] (mag arcsec$^{-2}$) | $r_e$ [b] kpc | Type | Redshift location | $V_{HI}$ km s$^{-1}$ | $W_{20}$ km s$^{-1}$ | $M_{HI}$ $\times 10^8$ M$_\odot$ |
|---|---|---|---|---|---|---|---|---|---|---|---|
| NGC 1200 | 8 (13464) | 3 3 13.5 | -11 55 56.1 | 0.56 | 24.4 | 2.3 | UDG | in group | 4047±6 | 34±6 | 1.8 |
| NGC 1200 | 23 (13548) | 3 4 16.7 | -12 3 51.0 | 0.47 | 23.3 | 4.0 | LSBG | background | 6856±6 | 85±6 | 7.1 |
| NGC 7396 | 5 (1817) | 22 50 24.0 | 1 11 56.4 | 0.27 | 23.7 | 2.1 | LSBG | background | 7360±6 | 62± 6 | 0.2 |
| NGC 7396 | 12 (2012) | 22 52 55.6 | 1 35 15.5 | 0.56 | 23.8 | 1.7 | LSBG | foreground | 3051±6 | 39±6 | 0.6 |
| NGC 7396 | 13 (2016) | 22 53 25.8 | 1 24 49.2 | 0.48 | 24.0 | 2.0 | UDG | in group | 4909±6 | 78±19 | 2.2 |

[a] g – band Central Surface Brightness from Tanoglidis et al. (2021) Catalogue version 1
[b] g – band effective radius estimated from Tanoglidis et al. (2021) Catalogue version 1 and detected H I velocities.
[c] Tanoglidis et al. (2021) Catalogue version 1 identifier numbers are provided in brackets

Mega-Science, Chinese Academy of Sciences, for the FAST distinguished young researcher fellowship (19-FAST-02) and the support from the National Natural Science Foundation of China (NSFC) under grant No. 12050410259 and the Ministry of Science and Technology (MOST) of China grant no. QNJ2021061003L. YZM acknowledges the support from National Research Foundation of South Africa with Grant No. 150580, No. 159044, No. CHN22111069370 and No. ERC23040389081. PL gratefully acknowledges support by the GEMINI ANID project No. 32240002. RK acknowledges the support of the Department of Atomic Energy, Government of India, under project no. 12-R&D-TFR-5.02-0700 and from the SERB Women Excellence Award WEA/2021/000008. DT acknowledges financial support from the XJTLU Research Development Fund (RDF) grant with number RDF-22-02-068. Part of the data published here have been reduced using the CARACal pipeline, partially supported by ERC Starting grant number 679627 "FORNAX", MAECI Grant Number ZA18GR02, DST-NRF Grant Number 113121 as part of the ISARP Joint Research Scheme, and BMBF project 05A17PC2 for D-MeerKAT. Information about CARACal can be obtained online under the URL: https://caracal.readthedocs.io. This research has made use of the NASA/IPAC Extragalactic Database (NED) which is operated by the Jet Propulsion Laboratory, California Institute of Technology, under contract with the National Aeronautics and Space Administration. This research has made use of the Sloan Digital Sky Survey (SDSS), http://www.sdss.org/.This letter also used CARTA software (Comrie et al. 2018).

REFERENCES

Abbott, T. M. C., Abdalla, F. B., Allam, S., et al. 2018, ApJS, 239, 18, doi: 10.3847/1538-4365/aae9f0

Blue Bird, J., Davis, J., Luber, N., et al. 2020, MNRAS, 492, 153, doi: 10.1093/mnras/stz3357

CASA Team, Bean, B., Bhatnagar, S., et al. 2022, PASP, 134, 114501, doi: 10.1088/1538-3873/ac9642

Comrie, A., Wang, K.-S., Hwang, Y.-H., et al. 2018, CARTA: The Cube Analysis and Rendering Tool for Astronomy, Zenodo, doi: 10.5281/zenodo.3377984

Crook, A. C., Huchra, J. P., Martimbeau, N., et al. 2007, ApJ, 655, 790, doi: 10.1086/510201

DESI Collaboration, Adame, A. G., Aguilar, J., et al. 2023, arXiv e-prints, arXiv:2306.06308, doi: 10.48550/arXiv.2306.06308

Díaz-Giménez, E., & Zandivarez, A. 2015, A&A, 578, A61, doi: 10.1051/0004-6361/201425267

For, B. Q., Spekkens, K., Staveley-Smith, L., et al. 2023, MNRAS, 526, 3130, doi: 10.1093/mnras/stad2921

Haynes, M. P., & Giovanelli, R. 1984, AJ, 89, 758, doi: 10.1086/113573

Hoosain, M., Blyth, S.-L., Skelton, R. E., et al. 2024, MNRAS, 528, 4139, doi: 10.1093/mnras/stae174

Józsa, G. I. G., White, S. V., Thorat, K., et al. 2020, in Astronomical Society of the Pacific Conference Series, Vol. 527, Astronomical Data Analysis Software and Systems XXIX, ed. R. Pizzo, E. R. Deul, J. D. Mol, J. de Plaa, & H. Verkouter, 635, doi: 10.48550/arXiv.2006.02955

Karunakaran, A., Spekkens, K., Zaritsky, D., et al. 2020, ApJ, 902, 39, doi: 10.3847/1538-4357/abb464

Mauch, T., Cotton, W. D., Condon, J. J., et al. 2020, ApJ, 888, 61, doi: 10.3847/1538-4357/ab5d2d

Meyer, M. J., Zwaan, M. A., Webster, R. L., et al. 2004, MNRAS, 350, 1195, doi: 10.1111/j.1365-2966.2004.07710.x

Palumbo, G. G. C., Saracco, P., Mendes de Oliveira, C., et al. 1993, ApJ, 405, 413, doi: 10.1086/172373

Pisano, D. J., Barnes, D. G., Staveley-Smith, L., et al. 2011, ApJS, 197, 28, doi: 10.1088/0067-0049/197/2/28

Plionis, M., Tovmassian, H. M., & Andernach, H. 2009, MNRAS, 395, 2, doi: 10.1111/j.1365-2966.2009.14507.x

Scott, T. C., Brinks, E., Cortese, L., Boselli, A., & Bravo-Alfaro, H. 2018, MNRAS, 475, 4648, doi: 10.1093/mnras/sty063

Sengupta, C., & Balasubramanyam, R. 2006, MNRAS, 369, 360, doi: 10.1111/j.1365-2966.2006.10307.x

Tanoglidis, D., Drlica-Wagner, A., Wei, K., et al. 2021, ApJS, 252, 18, doi: 10.3847/1538-4365/abca89

van der Burg, R. F. J., Hoekstra, H., Muzzin, A., et al. 2017, A&A, 607, A79, doi: 10.1051/0004-6361/201731335

van Dokkum, P. G., Abraham, R., Merritt, A., et al. 2015, ApJL, 798, L45, doi: 10.1088/2041-8205/798/2/L45

Xu, W., Ramos-Ceja, M. E., Pacaud, F., Reiprich, T. H., & Erben, T. 2018, A&A, 619, A162, doi: 10.1051/0004-6361/201833062

Zhou, Y.-F., Sengupta, C., Chandola, Y., et al. 2022, MNRAS, 516, 1781, doi: 10.1093/mnras/stac2344